\documentstyle[multicol,aps,pra]{revtex}

\draft
\begin{document}

\title{Fidelity for displaced squeezed thermal states 
and the oscillator semigroup}

\author{H. Scutaru}
\address{Department of Theoretical Physics, Institute of Atomic Physics\\
POB MG-6, Bucharest-Magurele, Romania\\
e-mail: {\rm scutaru@theor1.ifa.ro}}
\date{May 5, 1997}

\maketitle

\begin{abstract}
The fidelity for two displaced squeezed thermal states 
is computed using the fact that the corresponding density 
operators belong to the oscillator semigroup. 
\end {abstract}


\begin{multicols}{2}

The concept of fidelity is a basic ingredient in quantum
communication theory [1, 2]. Recently the corresponding
Bures distance was used [3] to define a measure of the
entanglement as the minimal Bures distance of an entangled
state to the set of disentangled states.

Let $\rho_{1}$ and $\rho_{2}$ be the density operators
which describe two impure states. The natural candidate for
the fidelity, denoted by $F(\rho_{1}, \rho_{2})$, is the
transition probability which must satisfy the following
natural axioms:
 
\begin{itemize}
{\small
\item {\bf F1} $F(\rho_{1}, \rho_{2}) \leq 1$ and
$F(\rho_{1}, \rho_{2})=1$ if and only if $\rho_{1} = \rho_{2}$;
\item {\bf F2} $F(\rho_{1}, \rho_{2}) = F(\rho_{2}, \rho_{1})$;
\item {\bf F3} If $\rho_{1}$ is a pure state $\rho_{1}=|\psi_{1}><\psi_{1}|$
then $F(\rho_{1}, \rho_{2})=<\psi_{1}|\rho_{2}|\psi_{1}>$;
\item {\bf F4} $F(\rho_{1}, \rho_{2})$ is invariant under unitary
transformations on the state space.}
\end{itemize}

Uhlmann's transition probability for mixed states [4]
\begin{equation}
F(\rho_{1}, \rho_{2})= [trace(\sqrt{\rho_{1}}\rho_{2}\sqrt{\rho_{1}})^{
{1 \over 2}}]^2
\end{equation}
does satisfy {\bf F1-F4}.

Investigations into detalied structure of the fidelity has been
hampered by the complicated square-root factors in (1). Due to these
technical difficulties in computing fidelity few concrete
results concerning the details of the fidelity have been found.
Until recently all known concrete results have been calculated
only for finite dimensional Hilbert spaces [5-7]. The first
result in an infinite dimensional Hilbert space has been obtained
by Twamley [8] for the fidelity of two undisplaced thermal states.
Twamley combines the Schur factorization with Baker-Campbell-
Hausdorff identities. But as he said these arguments do not
seem to hold for displaced squeezed thermal states. 
We have obtained the fidelity for two displaced
thermal states [9] using a result of Wilcox [10]. 

In the following we shall use a completely new method for the
treatement of the complicated square-root factors in (1)
in the case of two displaced squeezed thermal states (i.e.
in the case of two displaced mixed quasi-free states [11]).
The basic ingredient of the method is the oscillator semigroup
[12, 13]. The oscillator semigroup is the semigroup of
integral operators on $L^2({\bf R})$ whose integral kernels are 
Gaussians. The density operators which describe the displaced
squeezed thermal states belongs evidently to this semigroup.
Indeed the oscillator semigroup (or its closure) contains
the semigroup generated by the Hermite operator [13, 14] which is the
Hamiltonian of the quantum oscillator (and respectively
the range of metaplectic representation which generate the
squeezing [15, 16]).

The most general Gaussian density operator in coordinate
representation is an integral operator: 

\begin{equation}
(\rho \psi)(x) = \int_{-\infty}^{+\infty} <x|\rho|y> \psi(y) dy
\end{equation}
where
\begin{equation}
<x|\rho|y> = \exp{[-(ax^2+dy^2+2bxy)+lx+ky+g]}
\end{equation} 
In order that $\rho$ be a quantum density operator it must be
Hermitian, normalizable and non-negative [16]. Hermiticity for
$\rho$ requires $d=\bar a$, $b=\bar b$, $k= \bar l$ and $g=\bar
g$ [16]. From $trace \rho =1$ it follows that $g=-{(Re l)^2 \over
2(Re a+b)} - \ln{\sqrt{{ \pi \over 2(Re a +b)}}}$ and $Re a \geq
-b$. From the non-negativity of $<\psi|\rho|\psi>$ for all $|\psi>$
it follows [16] that $-b \geq 0$. Hence $Re a \geq -b \geq 0$.
For two quantum density operators $\rho_{1}$ and $\rho_{2}$ we
have the semigroup composition law
\begin{equation}
<x|\rho_{1}\rho_{2}|y> = \int_{-\infty}^{+\infty} <x|\rho_{1}|z> 
<z|\rho_{2}|y> dz
\end{equation}
Let us denote by $A$, $B$, $D$, $L$, $K$ and $G$ the
corresponding parameters of the Gaussian $<x|\rho_{1}\rho_{2}|y>$.
Then the semigroup composition law (the rule {\bf R1}) is given by

\begin{eqnarray}
&&
A = a_{1}-{b_{1}^2 \over d_{1}+a_{2}},~~D=d_{2}- {b_{2}^2 \over d_{1}+a_{2}}
\nonumber \\
&&
B=-{b_{1}b_{2} \over d_{1}+a_{2}},~~L = l_{1} - {(k_{1}+l_{2})b_{1} 
\over d_{1}+a_{2}}
\nonumber \\
&&
K= k_{2} - {(k_{1}+l_{2})b_{2} \over d_{1}+a_{2}}
\nonumber \\
&&
G=g_{1}+g_{2}+ {(k_{1}+l_{2}) \over 4(d_{1}+a_{2})}+
\ln{\sqrt{{\pi \over d_{1}+a_{2}}}}
\end{eqnarray}
We shall define the operator $\sqrt{\rho}$ as the integral operator
with the Gaussian kernel
\begin{equation}
<x|\sqrt{\rho}|y> = \exp{[-(\tilde{a}x^2+\tilde{d}y^2
+2\tilde{b}xy)+\tilde{l}x+\tilde{k}y+\tilde{g}]}
\end{equation}
such that
\begin{equation}
<x|\rho|y> = \int_{-\infty}^{+\infty} <x|\sqrt{\rho}|z> 
<z|\sqrt{\rho}|y> dz
\end{equation}
Then using the above given semigroup composition rule {\bf R1}
one obtains by inversion the following rule ({\bf R2}) for the
square-root of a density operator:
\begin{eqnarray}
&&
\tilde{a}=a-b
\nonumber \\
&&
\tilde{d}=d-b
\nonumber \\
&& 
\tilde{b}=-\sqrt{-b(a+d-2b)}
\end{eqnarray}
where the minus sign was choosen in order to have $-\tilde{b}
\geq 0$.
Also
\begin{eqnarray}
&&
\tilde{l}+\tilde{k}={l+k \over 1+2\sqrt{{-b \over a+d-2b}}}
\nonumber \\
&&
\tilde{l}-\tilde{k}=l-k
\nonumber \\
&& 
\tilde{g}={1 \over 2}g-{1 \over 2}\ln{\sqrt{{\pi \over a+d-2b}}}-
\nonumber \\
&&
{(l+k)^2 \over 8(\sqrt{a+d-2b}+2\sqrt{-b})^2} 
\end{eqnarray}
The characteristic function of the density operator $\rho$
is given by [11, 16]
\begin{eqnarray}
&&
CF_{\rho}(\alpha,\tau)= trace(W(\alpha,\tau)\rho)
=
\nonumber\\
&&
\exp{[-{1 \over 4}(a_{qq}\alpha^2+
a_{pp}\tau^2+2a_{pq}\alpha\tau)]}
\end{eqnarray}
where
\begin{eqnarray}
&&
(W(\alpha,\tau)\psi)(x)=\exp{[i\tau(x-{\alpha \over 2})]}
\psi(x-\alpha)
\nonumber \\
&&
a_{qq}=2(<Q^2>-<Q>^2)={1 \over a+d+2b}
\nonumber \\
&& 
a_{pp}=2(<P^2>-<P>^2)={4(ad-b^2) \over a+d+2b}
\nonumber \\
&&
a_{pq}=2<{1 \over 2}<QP+PQ>-<Q><P>)=
\nonumber\\
&&
{i(a-d) \over a+d+2b}
\end{eqnarray}
Here $(Q\psi)(x)=x\psi(x)$ and $(P\psi)(x)=-i{d\psi(x) \over dx}$.
\newline If $A$ is the matrix $\left(\matrix{a_{qq}&a_{pq} \cr a_{pq}&a_{pp}
\cr}\right)$ and $detA=a_{qq}a_{pp}-a_{pq}^2$ it is easy to
show that
\begin{equation}
a={detA+1 \over 4a_{qq}}-{ia_{pq} \over 2a_{qq}}
\end{equation}
$d=\bar a $ and
\begin{equation}
b=-{detA-1 \over 4a_{qq}}
\end{equation}
In order to simplify the calculations we shall use the property
{\bf F4}. Then it suffices to consider that $\rho_{1}$ is a
thermal state (i.e. an undisplaced and unsqueezed state) and that
only $\rho_{2}$ is a displaced squeezed thermal state. It is well known
that in this case [14]:
\begin{eqnarray}
&&
<x|\rho_{1}|y>=exp[-{1 \over 2}coth\beta~(x^2+y^2)+
\nonumber \\
&&
{xy \over sinh\beta}-\ln{ \sqrt{{\pi \over tanh\beta}}}]
\end{eqnarray}
If $\rho_{2}^{'}$ is an undisplaced squeezed thermal state
with:
\begin{equation}
<x|\rho_{2}^{'}|y> = \exp{[-(ax^2+dy^2+2bxy)+g^{'}]}
\end{equation}
then the displaced squeezed thermal state $\rho_{2}$ is
obtained as $\rho_{2}=W(\alpha,\tau)\rho_{2}^{'}W(-\alpha,-
\tau)$ and the corresponding kernel is given by
$<x|\rho_{2}|y>=\exp{[i\tau(x-y)]}<x-\alpha|\rho_{2}^{'}|x-\alpha>$
i.e.
\begin{equation}
<x|\rho_{2}|y> = \exp{[-(ax^2+dy^2+2bxy)+lx+ky+g]}
\end{equation} 
where $l=2(a+d)\alpha+i\tau$, $k=\bar l$ and $g=g^{'}-
(a+d+2b)\alpha^2$.
Now we can use the rules {\bf R1} and {\bf R2}. After
long but simple calculations we obtain the main result
of the paper:
\begin{eqnarray}
F(\rho_{1},\rho_{2})=
{2 \over \sqrt{\Delta+T}-\sqrt{T}}
\nonumber \\
&&
\exp{[-u^T(A_{1}+A_{2})^{-1}u]} 
\end{eqnarray}
where
$\Delta = det(A_{1}+ A_{2})$, $T=(detA_{1}-1)(detA_{2}-1)$ 
and where $u$ is the
column vector $\left(\matrix{\alpha \cr \tau \cr}\right)$.
A Gaussian density matrix describes a pure state if and only if
$detA=1$ [11]. If $\rho_{1}$ is a pure state then [11]:
\begin{eqnarray}
&&
F(\rho_{1},\rho_{2})=
trace\rho_{1}\rho_{2}=
\nonumber \\
&&
(2 \pi)^{-1}\int_{-\infty}^{+\infty}CF_{\rho_{1}}(-\alpha,
-\tau)CF_{rho_{2}}(\alpha,\tau)d \alpha d \tau 
\end{eqnarray}
and we obtain directly the same result as that obtained 
from the above formula:
\begin{eqnarray}
&&
F(\rho_{1},\rho_{2})=
\nonumber \\
{1 \over \sqrt{det({A_{1}+A_{2} \over 2})}}
\exp{[-u^T(A_{1}+A_{2})^{-1}u]}
\end{eqnarray}
The result of Twamley is reobtained for $u=0$ in a 
more compact form which is independent of the
parametrization. We remark that due to {\bf F4} the
formula (17) is generaly valid i.e. for any two displaced
squeezed thermal states. In order to compare our result
with that of Twamley [8] we shall use the canonical decomposition
of any correlation matrix (i.e. of any positive definite matix) 
$A$ (obtained with the aid of a theorem of
Balian, de Dominicis and Itzykson [17] concerning the canonical 
decomposition of symplectic matrices) (see also Ref. [11]):
\begin{equation}
A=O^T M \Gamma M O
\end{equation}
where $O=\left(\matrix{c&-s \cr s&~c \cr}\right)$,
with $c=cos\theta$ and $s=sin\theta$,
$M=\left(\matrix{m& 0 \cr 0& {1 \over m} \cr}\right)$
and $\Gamma=\left(\matrix{\gamma& 0 \cr 0& \gamma \cr}\right)$.
Then $detA=\gamma^2$ and
\begin{eqnarray}
&&
\Delta+T=\gamma_{1}^2\gamma_{2}^2+1+\gamma_{1}\gamma_{2}
[S^2((m_{1}m_{2})^2+{1 \over (m_{1}m_{2})^2})+
\nonumber \\
&&
C^2(({m_{1} \over m_{2}})^2+({m_{2} \over m_{1}})^2)]
\end{eqnarray}
where $C=cos(\theta_{2}-\theta_{1})$ and
$S=sin(\theta_{2}-\theta_{1})$.
The correspondence between the parametrization
from [8] and our parametrization is given by
$coshr={1 \over 2}(m+{1 \over m})$ and $cosh{\beta
\over 4}={\gamma \over \sqrt{\gamma^2-1}}$.
Now we consider the exponential factor
${\cal F}=\exp{[-u^T(A_{1}+A_{2})^{-1}u]}$.
The first remark concerns the following
form of $(A_{1}+A_{2})^{-1}$:
\begin{eqnarray}
(A_{1}+A_{2})^{-1}
=O_{1}^TM_{1}^{-1}
(\Gamma_{1}+\tilde \Gamma_{2})^{-1}
M_{1}^{-1}O_{1}
\end{eqnarray}
where
$\tilde \Gamma_{2}=
M_{1}^{-1}O_{1}O_{2}^TM_{2}\Gamma_{2}M_{2}O_{2}O_{1}^TM_{1}^{-1}$.
Then 
\begin{equation}
{\cal F}= \exp{[-\tilde{
u}^T(\Gamma_{1}+\tilde{\Gamma_{2}})^{-1}\tilde{u}]}
\end{equation}
where $\tilde u=M_{1}^{-1}O_{1}u$.
The matrix elements of ${\cal G}=(\Gamma_{1}+\tilde{\Gamma_{2}})^{-1}$
are given by
\begin{eqnarray}
&&
{\cal G}_{\tilde \alpha \tilde \alpha}={\gamma_{1}+
\gamma_{2}(S^2(m_{1}m_{2})^2+C^2({m_{1} \over m_{2}})^2)
\over \Delta}
\nonumber \\
&&
{\cal G}_{\tilde \tau \tilde \tau}={\gamma_{1}+
\gamma_{2}({S^2 \over (m_{1}m_{2})^2}+C^2({m_{2} \over m_{1}})^2)
\over \Delta}
\nonumber \\
&&
{\cal G}_{\tilde \alpha \tilde \tau}=-{\gamma_{2}
CS(m_{2}^2-{1 \over m_{2}^2})
\over \Delta}
\end{eqnarray}
When $\theta_{1}=\theta_{2}$ and $m_{1}=m_{2}=1$ we reobtain
the result from [9].

\vfill
\end{multicols}
\end{document}